\documentclass[prl,twocolumn,showpacs,groupedaddress]{revtex4}

\usepackage{dcolumn}
\usepackage{graphicx}
\usepackage{amsmath}
\usepackage{bm}
\usepackage[amssymb]{SIunits}
\usepackage{pdfpages}

\begin{document}

\title{Photon Blockade in the Ultrastrong Coupling Regime}

\author{A. Ridolfo$^1$, M. Leib$^1$, S. Savasta$^2$, and M. J. Hartmann$^1$}
\affiliation{$^1$Physik Department, Technische Universit\"{a}t M\"{u}nchen, James-Franck-Strasse, 85748 Garching, Germany\\
             $^2$Dipartimento di Fisica della Materia e Ingegneria Elettronica, Universit\`{a} di Messina Salita Sperone 31, I-98166 Messina, Italy} 
\date{\today}
\begin{abstract}
We explore photon coincidence counting statistics in the ultrastrong-coupling regime where the atom-cavity coupling rate becomes comparable to the cavity resonance frequency. In this regime usual normal order correlation functions fail to describe the output photon statistics.
By expressing the electric-field operator in the cavity-emitter dressed basis we are able to propose correlation functions that are valid for arbitrary degrees of light-matter interaction. Our results show that the standard photon blockade scenario is significantly modified for ultrastrong coupling. We observe parametric processes even for two-level emitters and temporal oscillations of intensity correlation functions at a frequency given by the ultrastrong photon emitter coupling.
These effects can be traced back to the presence of two-photon cascade decays induced by counter-rotating interaction terms.
\end{abstract}
\maketitle

The quantum theory of photodetection and optical coherence as originally formulated by Glauber \cite{Glauber} is central to all of quantum optics and
has occupied a key role in understanding radiation-matter interactions. Photodetection is also common to  quantum-state engineering \cite{Bimbard} and quantum information protocols \cite{Knill}.
One of the most prominent effects that shows nonclassical behavior of a quantum emitter is a phenomenon known as photon blockade \cite{Imamoglu,Birnbaum,Dayan}, where a coherent excitation of a cavity coupled to a highly nonlinear quantum system, such as a single atom, quantum dot or superconducting qubit, generates an output train of single photons. This photon statistics for the cavity output can be investigated by measuring the intensity correlation function $g^{(2)}(\tau)$, which demonstrates the nonclassical character of the transmitted field.

Recently a new regime of cavity quantum electrodynamics (cavity QED) has been reached experimentally where the strength of the interaction between an emitter and the cavity photons becomes comparable to the transition frequency of the emitter or the resonance frequency of the cavity mode \cite{guenter09,Niemczyk,Todorov,Schwartz,Hoffman, Scalari}.
In this so called ultrastrong coupling regime \cite{De Liberato,Peropadre2010,Nataf,guenter09,Niemczyk,Casanova} the routinely invoked rotating wave approximation is no longer applicable. As a consequence the number of excitations in the cavity-emitter system is no longer conserved, even in the absence of drives and dissipation. 

In this letter we explore the photon statistics of the output fields of a driven cavity QED system where a two level system (TLS) interacts with one cavity mode in the ultrastrong coupling regime. Specifically we show that the standard photon blockade effect does no longer persist since ultrastrong emitter-photon couplings enable processes where a single photon that enters the cavity can generate two or more output photons. Whereas such parametric processes can of course occur for emitters with three or more levels, we here show their existence for a two level system. We find that the two Rabi-splitted resonances, which are often termed upper and lower polariton, exhibit very different photon statistics. While excitation of the lower energy peak provides output light which is both subpoissonian and antibunched, as in the standard regime, excitation of the higher energy peak 
results in the emission of superpoissionan and bunched light. The calculated resonance fluorescence spectrum shows that this result originates from the activation of second order nonlinear optical effects which are absent in usual two level atoms.
Such a process can only result from counter-rotating terms in the atom-cavity interaction Hamiltonian.

As a second main result we find that the time dependence of photon-photon correlations $g^{(2)}(\tau)$ differs significantly from the standard photon blockade scenario. For standard photon blockade, one finds $g^{(2)}(0) \ll 1$ and oscillations with the Rabi frequency $\Omega$ of the laser drive for $\tau > 0$ (provided $\Omega$ exceeds the loss rates). The latter emerge since a time $\sim \Omega^{-1}$ is needed to load the next photon into the nonlinear cavity once a photon has left. For the ultrastrong coupling regime, we find oscillations at a frequency of the order of the emitter-photon coupling $g \gg \Omega$, i.e. much faster oscillations. These emerge since upper and lower polariton excitations can not be generated independently.
    
Our findings can be measured in present day experiments \cite{guenter09,Niemczyk,Todorov,Schwartz,Hoffman, Scalari}. A particularly well suited technology for such an experiment are superconducting circuits \cite{wallraff04,Norireview} which have recently emerged as an exquisite platform for microwave on-chip quantum-optics experiments.
Even though single photon detectors in the microwave regime are under current development \cite{Romero, Chen}, Glauber's first and second order correlation functions have been measured \cite{menzel10,Bozyigit} using quadrature amplitude detectors and standard photon blockade at microwave frequencies \cite{Lang} has been observed.

\paragraph{Input-output relations -}
Applying Glauber's idea of photodetection, we here introduce a full quantum theory to study the photon blockade in the ultrastrong coupling regime. 
This requires a proper generalization of input-output theory  \cite{GardinerZoller}, since the standard relations would
for example predict an output photon flux that is proportional to the average number of cavity photons, i.e. $\langle a_{\rm out}^\dag(t) a_{\rm out}(t) \rangle \propto \langle a^\dag(t) a(t) \rangle$ for vacuum input.
Hence an incautious application of this standard procedure to the ultrastrong-coupling regime would predict an unphysical stream of output photons for a system in its ground state which contains a finite number of photons due to the counter-rotating terms in the Hamiltonian. To remedy this problem, a model which explicitly takes into account the colored
nature of the dissipation bath has been proposed \cite{Ciuti,DeLiberato}. Yet such a procedure is numerically quite demanding and would require the derivation (and also the time-integration) of four-time correlation functions for the calculation of $g^{(2)}(\tau)$.

Here we propose a different route going back to Glauber's formulation of photodetection \cite{Glauber}. By expressing the cavity electric-field operator in the atom-cavity dressed basis we derive correlation functions for the output fields which are valid for arbitrary degrees of light-matter interaction in the cavity. The probability per unit time that a photon be absorbed by an ideal detector is proportional to $\langle E^-(t) E^+(t) \rangle$ were $E^\pm(t)$ are the positive and negative frequency components of the electric field operator of the output field \cite{Glauber,Milonni} and higher order photon correlation funtions e.g. read $\langle E^-(t) E^-(t') E^+(t') E^+(t) \rangle$. In circuit QED the same quantities are measured via output voltages which are proportional to the electric fields.

To derive the appropriate input-output relations \cite{GardinerZoller}, we assume a cavity that is coupled to a one-dimensional output waveguide via an interaction between the cavity field $X$ and the momentum quadratures $\Pi_{\omega}$ of the waveguide field outside the cavity,
$H_{I} = \epsilon_{c} \int d\omega X \Pi_{\omega}$ with $\Pi_{\omega} = - i \sqrt{\frac{\hbar \omega}{4 \pi \epsilon_{o} v}} \left[a_{\omega}(t) - a_{\omega}^{\dagger}(t) \right]$,
where $\epsilon_{c}$ is a coupling parameter and $\epsilon_{o}$ is a parameter describing the dielectric properties of the output waveguide, 
$v$ is the phase velocity and $a_{\omega}$($a_{\omega}^{\dagger}$) annihilation(creation) operators of the fields outside the cavity.
This form of the coupling applies to optical implementations where the electric fields in- and outside the cavity couple as well as to circuit QED setups where voltages (and hence electric fields) of the cavity and output line couple via a capacitance. For optical realizations, $\epsilon_{o}$ is thus the permittivity and for circuit QED setups the capacitance per unit length of the output line. 
We define output (input) field operators,
\begin{equation} \label{eq:output-def}
a_{\text{out}(\text{in})}(t) = \frac{1}{\sqrt{2 \pi}} \int d\omega \sqrt{\omega} e^{-i \omega (t - \tilde{t})} a_{\omega}(\tilde{t}) ,
\end{equation}
where $\tilde{t} = t_{1} \to  +\infty$ for the output field and $\tilde{t} = t_{0} \to  -\infty$ for the input field.
With this definition, $\hbar \langle a_{\text{out}}^{\dagger}(t) a_{\text{out}}(t) \rangle$ is proportional to the measured $\langle E^-(t) E^+(t) \rangle$ and describes an energy flux associated to the output light. The definition of output fields as used in many textbooks, c.f. \cite{Walls}, is recovered if all frequencies of the field are very close to a ``carrier'' frequency $\overline{\omega}$ and one may approximate $\sqrt{\omega} \approx \sqrt{\overline{\omega}}$ in the integral kernel which makes the observed energy fluxes proportional to photon number fluxes.
Following the standard procedure we obtain the input-output relation,
\begin{equation} \label{eq:input-output}
a_{\text{out}}(t) = a_{\text{in}}(t) - i \frac{\epsilon_{c}}{\sqrt{8 \pi^{2} \hbar \epsilon_{o} v}} \dot{X}^{+},
\end{equation}
where we have applied a rotating wave approximation to $H_{I}$ since $\omega \gg |\epsilon_{c}/(\sqrt{8 \pi^{2} \hbar \epsilon_{o} v})|$. One thus needs to find the positive frequency component of $\dot{X}$ according to its actual dynamical behavior, c.f. \cite{Savasta96}, to compute the proper output fields. We do this by expressing $X$ in the atom-cavity dressed basis.
Importantly, in the ultrastrong coupling regime, the positive frequency component of $X$ is not proportional to the photon annihilation operator $a$.
We now turn to introduce our model.

\begin{figure}[!ht]  
\includegraphics[height=40mm]{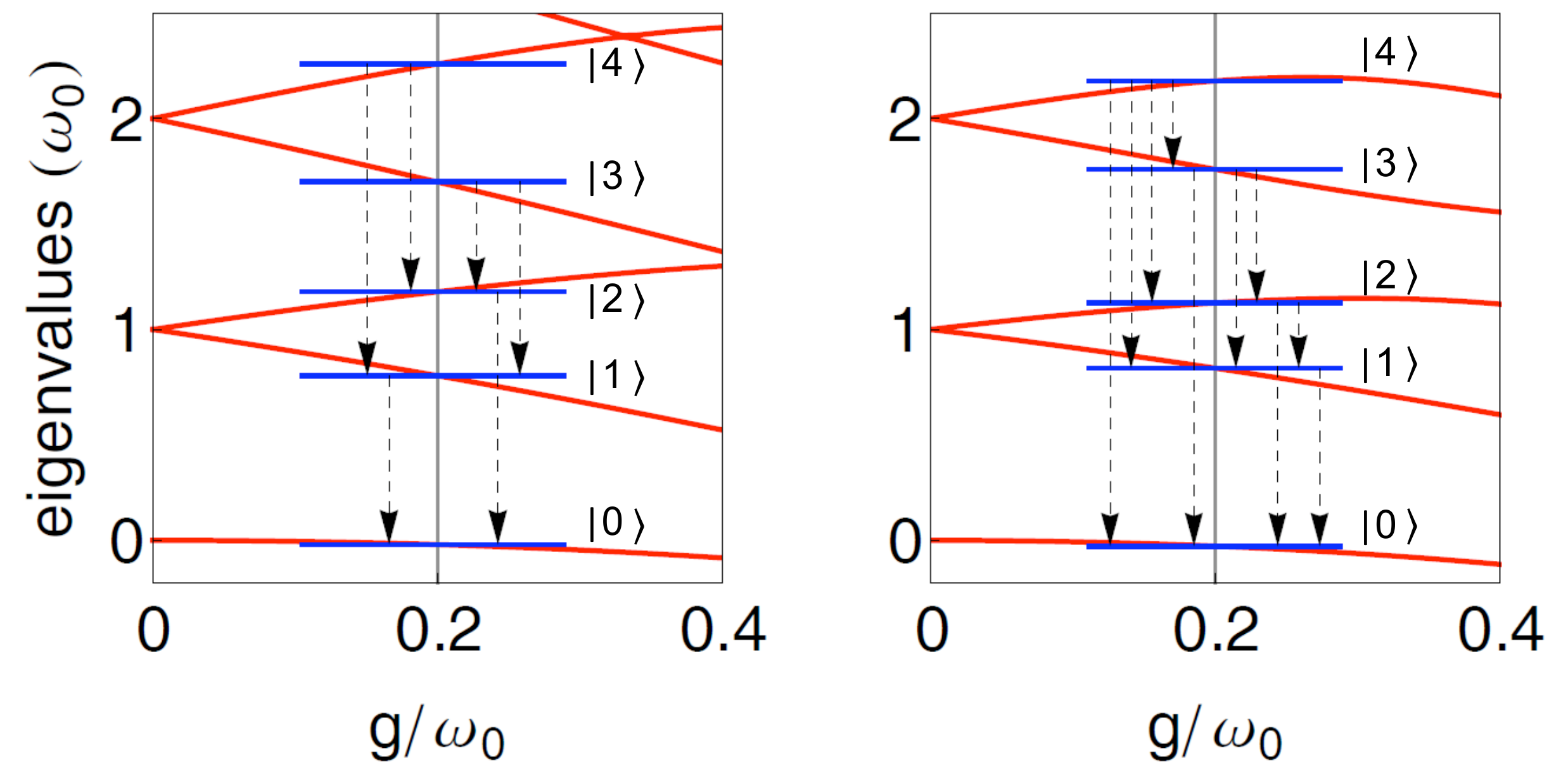}  
\caption{(color online) Energy spectrum of $H_{\rm 0}$ as function of the coupling strength for $\theta =  \pi/2$ (left panel) and $\theta =  0.93$ (right panel). In both panels $\omega_{\rm x} = \omega_{\rm 0} $. The vertical line marks the coupling strength ($g = $ 0.2 $\omega_{\rm 0}$) that is used in the sequel. For $\theta =  \pi/2$ the level structure is analogous to that of the JC model. The arrows indicate possible transitions of radiative decay.} 
\label{fig:spectrum}
\end{figure}

\paragraph{Model -}
We consider a coherently driven cavity QED system with the most general linear coupling between a single cavity mode and a two level system (TLS). Its Hamiltonian reads (we set $\hbar = 1$),
\begin{align} \label{eq:model}
       H_{\rm S} & = H_{\rm 0} + H_{\rm drive} \quad \text{with} \\
       H_{\rm 0} & = \omega_{\rm 0} a^{\dagger}a + \omega_{\rm x} \sigma^{+} \sigma^{-} + g ( a + a^{\dagger})\left[\cos(\theta)\sigma_{\rm z} - \sin(\theta)\sigma_{\rm x}\right] \nonumber
\end{align} 
Here, $H_{\rm 0}$ is the energy of the TLS described by the Pauli operators $\sigma_{\rm x,y,z}$ ($\sigma^{\pm} = (\sigma_{\rm x} \pm i \sigma_{\rm y})/2$) and the cavity mode with annihilation (creation) operators $a (a^{\dagger})$.  $g$ describes the strength and the mixing angle
$\theta$ the further properties of the coupling between the TLS and the cavity mode. An additional coupling term $\propto  ( a + a^{\dagger})\sigma_{\rm y}$ would not change the physics as it could be compensated by a rotation around the z-axis.  $H_{\rm drive} = \Omega \cos(\omega_{\rm d} t)(a + a^{\dagger})$ describes the coherent driving of the cavity mode with frequency $\omega_{\rm d}$ and drive amplitude $\Omega$. 
The mixing angle $\theta$ plays a crucial role since both, the spectrum of $H_{\rm 0}$ and the output photon statistics strongly depend on its value. 
In particular, for $\theta = \pi/2$ the Hamiltonian $H_{\rm 0}$ conserves the parity of the number of excitations in the system, whereas this is no longer the case for $\theta \neq \pi/2$ \cite{Deppe08,Niemczyk11,Liu}. The latter enables transitions that are forbidden in the usual Jaynes-Cummings (JC) model, c.f. Fig. \ref{fig:spectrum}, and leads to novel effects in the photon statistics. Note that for weaker couplings ($g \ll \omega_{\rm 0}, \omega_{\rm x}$) the term $g ( a + a^{\dagger})\cos(\theta)\sigma_{\rm z}$ can be neglected in a rotating wave approximation and parity is always conserved.

For describing a realistic system, dissipation induced by its coupling to the environment needs to be considered. Yet, owing to the very high ratio $g/\omega_{\rm 0}$,
a standard quantum optical master equations fails as it would for example predict that even zero temperature environments could drive the system out of its ground state.
A viable description of the system's coupling to its environment requires a perturbative expansion in the system-bath coupling strength.
To accurately perform this expansion we write the Hamiltonian in a basis formed by eigenstates $|j\rangle$ of $H_{0}$, denote the respective energy eigenvalues by $\hbar \omega_j$, i.e. $H_{0}|j\rangle = \hbar \omega_j |j\rangle$, and derive Redfield equations \cite{breuer} to describe the dissipative processes \cite{Blais}. We choose the labeling of the states $|j\rangle$ such that $\omega_k > \omega_j$ for $k > j$ and focus on a single-mode cavity and a $T=0$ temperature environment.
Generalizations to a multi-mode cavity and $T\neq0$ environments are straightforward. Assuming a weak laser drive, $\Omega \ll g, \omega_{\rm 0}, \omega_{\rm x}$, and treating $\Omega$ perturbatively to leading order, see supplementary information, we thus arrive at the master equation \footnote{We neglect small Lamb shifts as they do not alter the output photon statistics.},
\begin{equation}\label{master-eq}
    \dot\rho(t) = i [\rho(t), H_{\rm S}] + \mathcal{L}_{a}\rho(t) + \mathcal{L}_{x}\rho(t).
\end{equation}
Here $\mathcal{L}_{a}$ and $\mathcal{L}_{x}$ are Liouvillian superoperators describing the losses of the system where
$ \mathcal{L}_{c}\rho(t) = \sum_{j,k>j}\Gamma^{j k}_{c} \mathcal{D}[|j \rangle \langle k|]\rho(t)$ for $c = a, \sigma^{-}$ and $\mathcal{D}[\mathcal{O}]\rho = \frac{1}{2} (2 \mathcal{O}\rho\mathcal{O}^{\dagger}-\rho \mathcal{O}^{\dagger} \mathcal{O} - \mathcal{O}^{\dagger} \mathcal{O}\rho)$. Standard dissipators are recovered in the limit $g \to 0$.
The relaxation coefficients $\Gamma^{j k}_{c} = 2\pi d_{c}(\Delta_{k j}) \alpha^{2}_{c}(\Delta_{k j})| C^{c}_{j k}|^2$ depend on the spectral density of the baths $d_{c}(\Delta_{k j})$ and the system-bath coupling strength $\alpha_{c}(\Delta_{k j})$ at the respective transition frequency $\Delta_{k j} = \omega_{k} - \omega_{j}$ as well as on the transition coefficients $C_{j k} = -i \langle j |(c - c^{\dagger})| k \rangle$
($c = a, \sigma^{-}$). These relaxation coefficients can be interpreted as the full width at half maximum of each $|k \rangle\rightarrow | j \rangle$ transition. As we consider a cavity that couples to the momentum quadratures of fields in one-dimensional output waveguides, we assume the spectral densities $d_{c}(\Delta_{k j})$ to be constant and $\alpha_{c}^{2}(\Delta_{k j}) \propto \Delta_{k j}$. Hence the relaxation coefficients reduce to $ \Gamma^{j k}_{c} = \gamma_{c} \,\frac{\Delta_{k j}}{\omega_{0}} \, |C^{c}_{j k}|^2$, where $\gamma_{c}$ are the standard damping rates of a weak coupling scenario. In equation (\ref{master-eq}) contributions of dephasing noise were omitted as these only lead to  very minor quantitative modifications of our findings, see supplementary information.

\paragraph{Results -}
According to the input-output relation (\ref{eq:input-output}) the photon statistics of the output light is, for input fields in vacuum, equal to,
\begin{equation} \label{eq:g2ofX}
g^{(2)}(\tau) = \lim_{t \to \infty} \frac{\langle \dot{X}^{-}(t) \dot{X}^{-}(t+\tau) \dot{X}^{+}(t+\tau) \dot{X}^{+}(t) \rangle}{\langle \dot{X}^{-}(t) \dot{X}^{+}(t) \rangle^{2}},
\end{equation}
where $X = -i X_{0} (a - a^{\dag})$. We thus compute $g^{(2)}(\tau)$ from the stationary state solution of Eq. (\ref{master-eq}). To separate $\dot{X}$ in its positive and negative frequency components, $\dot{X}^{+}$ and $\dot{X}^{-}$, we expand it in terms of the energy eigenstates $|j\rangle$ and find $\dot{X}^+ = -i \sum_{j,k>j} \Delta_{kj}X_{jk} | j \rangle \langle k |$, where $X_{jk} = \langle j | X | k \rangle$ and $X^- = (X^+)^\dag$. Note that $X^+ |0 \rangle = 0$, for the system ground state $|0 \rangle$ in contrast to $a |0 \rangle \neq 0$.

\begin{figure}[!t]  
\includegraphics[height=40mm]{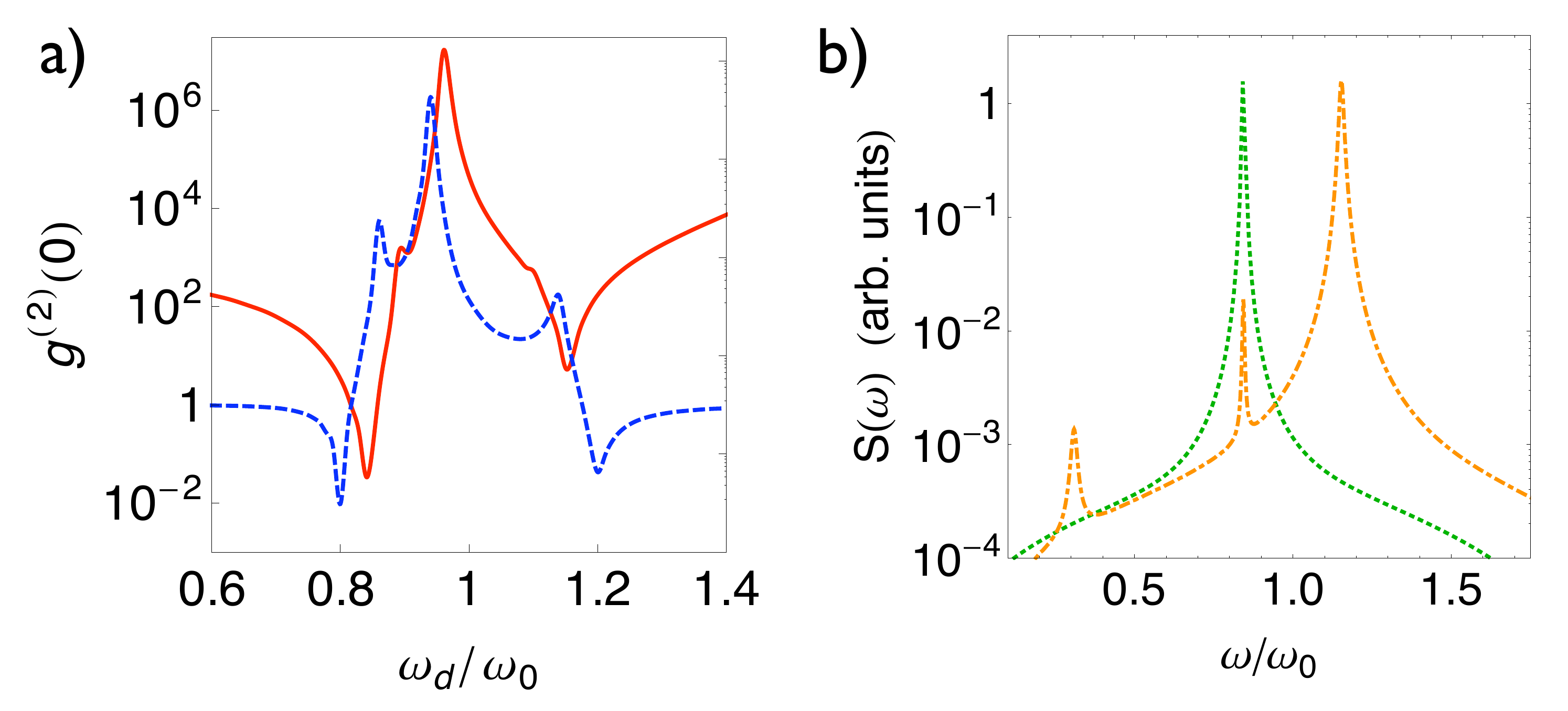}  
\caption{(color online) Emission characteristics of our system. {\bf a)} $g^{(2)}(0)$ as function of the frequency, $\omega_{\text{d}}$, of the coherent drive for $\theta = 0.93$ (red continuous line) and $\theta = \pi/2$ (blue dashed line). 
{\bf b)} Resonance fluorescence spectra calculated for a driving field tuned with the transition $|0 \rangle \to |1 \rangle$, i.e. the lower polariton, (green dashed line) and with the transition $|0 \rangle \to |2 \rangle$, i.e. the upper polariton, (orange dash-dotted line) for $\theta = 0.93$.
The remaining parameters are $\omega_{\rm x} = \omega_{\rm 0}$, $\gamma_{a} = \gamma_{x} = 10^{-2}$ $\omega_{\rm 0}$, $\Omega = 10^{-4}$ $\omega_{\rm 0}$, $g = 0.2$ $\omega_{\rm 0}$ for all plots.} 
\label{fig:g2(0)}
\end{figure}

Fig. \ref{fig:g2(0)} a), shows $g^{(2)}(\tau = 0)$ for $\omega_{\rm x} = \omega_{\rm 0}$, $\gamma_{a} = \gamma_{x} = 10^{-2}$ $\omega_{\rm 0}$, $\Omega = 10^{-4}$ $\omega_{\rm 0}$, $g = 0.2$ $\omega_{\rm 0}$.
The blue-dashed line represents $g^{(2)}(0)$ for the parity conserving model ($\theta = \pi/2$), where a suppression of the probability to measure two photons per count is evident for the two polaritonic peaks. This has the standard explanation that only one photon at a time can be absorbed on the respective transitions due to the nonlinearity of the system.
Instead, the strong bunching between both polaritonic peaks ($\omega_{\text{d}} \sim \omega_{0}$) is due to the simultaneous excitation of both polaritons. By varying $\theta$ we observe a new and anomalous effect. For $\theta = 0.93$ the expected anti-bunching for the higher frequency dip
($\omega_{\text{d}} \approx 1.18 \omega_{0}$) disappears and turns into a pronounced bunching ($g^{(2)}(0) \sim 5.6$),
c.f. Fig.\ref{fig:g2(0)} a). We find a similar behavior for lower values of $\theta$. 

An explanation of this exotic behavior can be found from the dynamics of the decay and excitation mechanisms, where it is crucial which transitions between energy levels exhibit radiative decay. To probe the latter we calculated the incoherent part of the scattered light, i.e. the resonance fluorescence spectrum which reads
\begin{equation}
S(\omega) \propto \lim_{t \to \infty} 2 \Re \int_{0}^{\infty} \langle \dot{X}^{-}(t)\dot{X}^{+}(t+\tau)\rangle e^{i \omega \tau} d\tau
\end{equation}
for input fields in vacuum. Here $\Re$ denotes the real part. For a weak excitation density $S(\omega)$ is proportional to the emission.
Fig. \ref{fig:g2(0)} b) shows the resonance fluorescence spectrum for a weak driving field resonant with the transitions $|0 \rangle \to |1 \rangle$ (blue dashed line) and $|0 \rangle \to |2 \rangle$ (red continuous line). The first case exhibits a single peak of typical Lorenzian shape reminiscent of a weakly driven TLS. Here, only level $|1\rangle$ is efficiently excited due to strong non-linearity
and hence emission predominantly occurs on the transition $|1 \rangle \to |0 \rangle$. Instead, when the system is resonantly excited on the $|0 \rangle \to |2 \rangle$ transition, the resonance fluorescence displays a triplet structure. These peaks arise from the $|2 \rangle \to |0 \rangle$, $|2 \rangle \to |1 \rangle$ and $|1 \rangle \to |0 \rangle$ transitions.
This triplet is a peculiarity of the interaction terms proportional to $\sigma_{\rm z}$ in $H_{\rm 0}$, for which the matrix elements $X_{jk}^{+}$ for all involved levels $j,k = 0,1,2$ are non-zero and thus enable parametric multi-photon processes with probabilities proportional to $|X_{jk}^{+}|^2$. Here we create an excitation in level 2 by resonantly driving the $|0 \rangle \to |2 \rangle$ transition and the system can decay via the cascade $|2 \rangle \to |1 \rangle \to |0 \rangle$ which gives rise to the bunching effect observed in $g^{(2)}(0)$, c.f. Fig. \ref{fig:g2(0)} a). Whenever $\theta = \pi/2$, the standard dipolar selection rules as in the JC model are recovered and only transitions between states that differ by one excitation number are allowed. Hence, there is no radiative decay on the $|2 \rangle \to |1 \rangle$ transition and $g^{(2)}(0)$ shows the characteristic antibunching even if we drive the system resonantly with $|0 \rangle \to |2 \rangle$ (blue dashed line in Fig. \ref{fig:g2(0)} a)).     

\begin{figure}[!ht]  
\centering
\includegraphics[height=36mm]{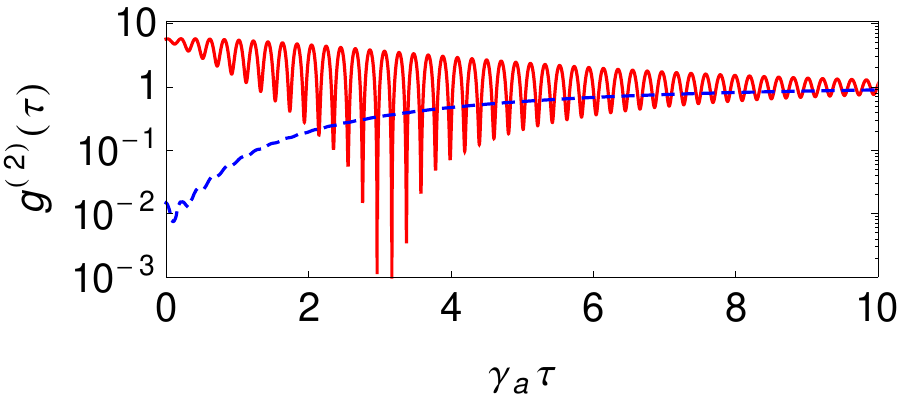}  
\caption{(color online) $g^{(2)}(\tau)$ for a driving frequency resonant to the energy of the first (blue dashed line) and on the second polariton (red continuous line). $\omega_{\rm x} = \omega_{\rm 0}$, $\gamma_{a} = \gamma_{x} = 10^{-2}$ $\omega_{\rm 0}$, $\Omega = 10^{-4}$ $\omega_{\rm 0}$, $g = 0.2$ $\omega_{\rm 0}$ and $\theta = 0.93$ for both cases.} 
\label{fig:g2(tau)}
\end{figure}
%
Moreover, driving the $|0 \rangle \to |2 \rangle$ transition triggers oscillations of $g^{(2)}(\tau)$ between bunching and antibunching values at a frequency $\Delta_{21}$ as shown in Fig. \ref{fig:g2(tau)}. Such oscillations also appear in a driven Jaynes-Cummings system albeit at much smaller amplitude but are absent for a single driven two level atom. Their appearance can be explained by the presence of different types of excitations.
For a two level atom there is only one type of excitations with one resonance frequency in the system. Hence all dynamics in $g^{(2)}(\tau)$ is due to loading and emission processes only. For a driven Jaynes-Cummings system, there are two types of excitations, the lower polaritons and the upper polaritons which differ in frequency by $\Delta_{21}\sim 2 g$ and one sees oscillations of this frequency in $g^{(2)}(\tau)$. Yet if one drives one of the Rabi peaks, only one polariton species is predominantly excited and the amplitude of the oscillations is very small. In contrast, if one drives the system of Eq. (\ref{eq:model}) in the ultrastrong coupling regime on the transition $|0 \rangle \to |2 \rangle$, one always significantly excites two excitations species whose frequencies differ by $\Delta_{21}$ which in turn causes oscillations of large amplitude in $g^{(2)}(\tau)$.
Oscillations of the same frequency also appear in $g^{(2)}(\tau)$ for $\omega_{\rm d} = \omega_{01}$, c.f.  Fig. \ref{fig:g2(tau)} (blue dashed line), but are much smaller in amplitude as the level $|3\rangle$ is barely excited. 
The appearance of these pronounced oscillations can thus also be traced back to parametric processes enabled by the ultrastrong coupling strength.

\paragraph{Circuit QED implementation --}
We finally discuss an experimental setup with superconducting circuits that is ideally suited for observing our findings \cite{Leib12}.
The Hamiltonian of Eq. (\ref{eq:model}) is realized in a coplanar waveguide that is galvanically coupled to a flux qubit \cite{Niemczyk}. In this implementation, the mixing angle $\theta$ is related to the bias flux through the qubit loop. 
For ultrastrong couplings a 
description with a TLS and a single cavity mode is a valid approximation provided the flux threaded through the qubit loop is chosen such that $\cos\theta \lesssim 0.48$, see supplementary information.

\paragraph{Conclusions --}
We have explored the output photon statistics of a driven cavity QED system that operates in the ultrastrong coupling regime.
We find that the prominent photon blockade does not persist in its known form for such strong couplings since parametric processes emerge.
For efficient coupling between cavity and output modes, the latter could be exploited for building high yield parametric down-converters \cite{Moon05}. Generalizations of our study to frequency resolved detection \cite{delValle12} and multi-cavity devices \cite{Leib10} would form interesting perspectives for future research.

This work is part of the Emmy Noether project HA 5593/1-1 and was supported by the CRC 631, both funded by the German Research Foundation, DFG.


\section{supplementary information}

\subsection{Perturbative treatment of the driving term}

A derivation of a quantum master equation leads to the equation \cite{GardinerZoller},
\begin{eqnarray} \label{eq:master1}
\dot{\rho}(t) & = & -i [H_{S}(t), \rho_{S}(t)] \\
 & - & \int_{t_{0}}^{t} ds \, \text{Tr}_{E} \left\{ \left[ H_{I}, \left[ H_{I}(s-t), \rho_{E} \otimes \rho(t) \right] \right] \right\} \nonumber ,
\end{eqnarray}
where $H_{I}(s-t) = U(t-s) H_{I} U^{\dagger} (t-s)$ with,
\begin{equation}
U(t-s) =  e^{-iH_{E}(t-s)} \, \mathcal{T}_{+} e^{-i \int_{0}^{t-s} d\tau H_{S}(\tau)} 
\end{equation}
Here $H_{S}(t)$ is the time dependent Hamiltonian of our system that includes the time dependent drive term, c.f. equation (3) of the main text, $\mathcal{T}_{+}$ is the appropriate time ordering operator and
$H_{E}$ is the Hamiltonian of the environment. In turn, $\rho(t)$ is the density matrix of the system and $\rho_{E}$ the state of the environment which is the vacuum in our case. Our perturbative treatment of the drive consists in approximating,
\begin{equation}
U(t-s) \approx  e^{-iH_{E}(t-s)} \, e^{-i H_{0}(t-s)} ,
\end{equation}
where $H_{0}$ is the system Hamiltonian without the drive term as in the main text.
For, $\Omega \ll g, \omega_{\rm 0}, \omega_{\rm x}$ this is a good approximation since the right hand side of equation (\ref{eq:master1}) is already 2nd order in the weak system environment coupling. Our approach is thus a second order expansion in both, the system environment coupling and the drive amplitude $\Omega$.
 
We furthermore apply a rotating wave approximation which discards all processes where system and environment respectively system and drive would both gain (loose) energy at the same time.

Due to the time dependence of $H_{S}(t)$, even the asymptotic state, $\lim_{t \to \infty} \rho (t)$, has small residual oscillations. For calculating steady state expectation values we average in time over several such residual oscillations.
This closely corresponds to the time integration of the output signal that is usually applied to record experimental data.

\subsection{Circuit QED implementation}

Experimental setups with superconducting circuits are ideally suited for observing our findings.
The Hamiltonian of Eq. (3) in the main text is realized in a transmission line resonator that is galvanically coupled to a flux qubit \cite{Niemczyk}. In this implementation, the mixing angle $\theta$ is related to the bias flux through the qubit loop. With respect to the experimental feasibility for measuring the effects we predict, it is important to verify the validity of the description with a two level system (TLS) and a single cavity mode. 
The nonlinearity of typically employed flux qubits \cite{Niemczyk} greatly exceeds the qubit transition frequency so that a TLS description is very accurate.
As we consider the lowest frequency mode of the cavity, couplings to higher modes can be neglected provided that processes that de-excite the qubit and destroy a photon in the lowest frequency mode to create a photon in the second harmonic mode are ineffective.
We now turn to estimate when this is the case.

The Hamiltonian for the setup comprising all resonator modes reads,
\begin{eqnarray}
H&=&\hbar \omega_q \frac{\sigma_z}{2}+\sum_n \hbar\omega_n (a_n^{\dag}a_n+\frac{1}{2})\\
 &+& \sum_{n} \hbar g_n (a_n^{\dag}+a_n)(\cos\theta\sigma_z-\sin\theta\sigma_x) \nonumber
\end{eqnarray}
with,
$ \cos\theta=\frac{2I_p\delta\phi}{\sqrt{\Delta^2+(2I_p\delta\phi)^{2}}}$
and
$\sin\theta=\frac{\Delta}{\sqrt{\Delta^2+(2I_p\delta\phi)^{2}}}$,
where $\omega_q$ and $\omega_n$ are the qubit and resonator frequencies respectively, $g_n$ is the qubit resonator-mode coupling and $\delta\phi$, $\Delta$ and $I_p$ are the flux threaded through the loop, the sweet spot frequency $\omega_q(\delta\phi=0)$ and the persistent circulating current of the flux qubit. Numerical values for all these parameters extracted from \cite{Niemczyk} are tabularized in table \ref{NiemTable}. The value of the qubit transition frequency as used in the main text is the value of $\omega_{q}$ for the chosen flux bias $\delta\phi_{0}$,
i.e. $\omega_{\rm X} = \omega_{q}(\delta\phi_{0})$.
\begin{table}
\caption{\label{NiemTable} }
\setlength{\extrarowheight}{10pt}
\begin{tabular}{r|l}
$\frac{\Delta}{h}$/$2I_p$  & $\unit{2.25}{\giga\hertz}$/$\unit{630}{\nano\ampere}$\\[5pt]
\hline
$\frac{\omega_1}{2\pi}$/$\frac{\omega_2}{2\pi}$/$\frac{\omega_3}{2\pi}$ & $\unit{2.782}{\giga\hertz}$/$\unit{5.357}{\giga\hertz}$/$\unit{7.777}{\giga\hertz}$\\[5pt]
\hline
$\frac{g_1}{2\pi}$/$\frac{g_2}{2\pi}$/$\frac{g_3}{2\pi}$ & $\unit{314}{\mega\hertz}$/$\unit{636}{\mega\hertz}$/$\unit{568}{\mega\hertz}$
\end{tabular}
\end{table}

\begin{figure}
\centering
\includegraphics[width=7cm]{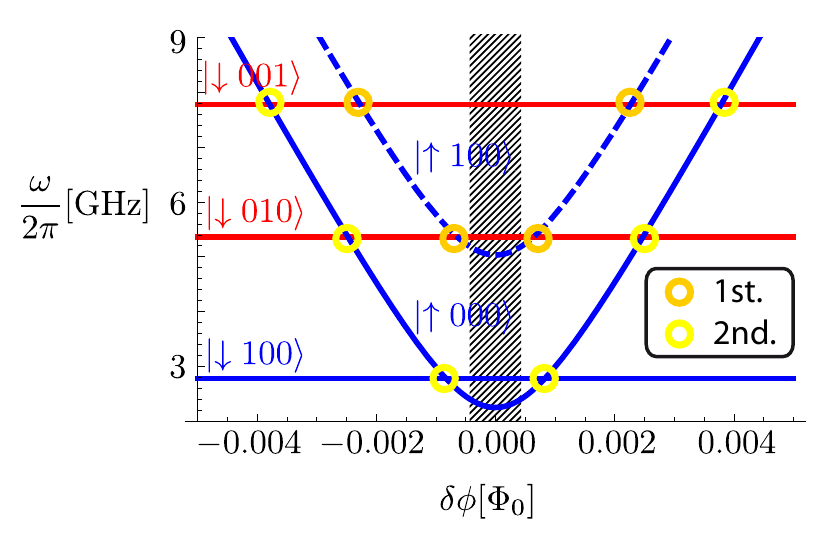}
\caption{Eigenenergies of the uncoupled resonator and flux qubit system as a function of the flux threaded through the qubit loop. The energies of states that belong to the Hilbert space of our description (TLS and lowest frequency resonator mode) are drawn in blue whereas states that do not belong to this Hilbert space are drawn in red. Anticrossings due to interaction are labeled by the order of perturbation theory that lifts the degeneracy. The crosshatched area marks the allowed range of $\delta\phi$.}
\label{spec}
\end{figure}

Our goal is to find parameter ranges where the embedded flux qubit despite the utrastrong coupling regime only couples to the fundamental, i.e. the lowest frequency resonator mode. 
To this end we consider the eigenstates of a noninteracting system of flux qubit and resonator which are direct products of the eigenstates of the qubit and the resonator, e.g. $\left|\uparrow \downarrow, n_1 n_2 n_3\right\rangle$ and have eigenenergies $E=\pm\hbar\omega_q+\sum_lÊ\hbar\omega_l (n_l+\frac{1}{2})$. Here, frequency shifts of the qubit and resonator due to their mutual presence are already taken into account, c.f. \cite{Leib12}. Experiments show that it suffices to take only the three lowest eigenmodes of the resonator into account \cite{Niemczyk}.
The qubit-resonator interaction couples these eigenstates. For the single-mode description of the cavity to be valid, couplings between states $\left|\uparrow \downarrow, n_1 0_2 0_3\right\rangle$ and states $\left|\uparrow \downarrow, 0_1 n_2 n_3\right\rangle$ with $n_{2} \ge 1$ or $n_{3} \ge 1$ need to be negligible. They can indeed be neglected provided the coupling is small compared to the difference of their energies.
 The eigenenergies of the noninteracting system are plotted as a function of the flux threaded through the qubit loop in figure \ref{spec}. We focus on the area around the sweet spot ($\delta\phi = 0$) and thus need to keep $\delta\phi$ small enough so that the energy difference between the state $\left|\uparrow 1_{1}0_{2}0_{3}\right\rangle$ and the state $\left|\downarrow 0_{1}1_{2}0_{3}\right\rangle$ stays large compared to their mutual coupling.
The energies of states $\left|\downarrow n_{1}0_{2}0_{3}\right\rangle$ with $n_{1} \ge 2$ have been omitted in figure \ref{spec} since they are either well separated from $\left|\downarrow 0_{1}1_{2}0_{3}\right\rangle$ or only cross $\left|\downarrow 0_{1}1_{2}0_{3}\right\rangle$ for much larger values of $\delta\phi$.

On resonance between $\left|\uparrow 1_{1}0_{2}0_{3}\right\rangle$ and $\left|\downarrow 0_{1}1_{2}0_{3}\right\rangle$
we find for their coupling, 
\begin{equation}
J=\hbar\frac{g_1g_2\cos\theta\sin\theta}{\omega_q+\omega_2}\,,
\end{equation} 
in second order perturbation theory in $g_{1}$ and $g_{2}$. Despite the ultrastrong coupling, perturbation theory yields a reliable estimate for the values of $g/\omega_{0} = 0.2$ as used in the main text.

We need to make sure that the energy difference between the two states always exceeds this coupling energy $J$. This restricts the allowed range of the flux threaded through the qubit loop and in turn $\cos\theta$ is bound from above by $\cos\theta_{max}\approx 0.48$ 

\subsection{Effect of qubit dephasing}

In typical circuit QED implementations of the ultrastrong coupling regime, qubit dephasing can not be neglected. To analyse its impact on our findings, we present here results for a system where dephasing takes place at the same rate as damping, i.e. $\gamma_{\text{deph}} = \gamma$. 

\begin{figure}  
\includegraphics[width=5.4cm]{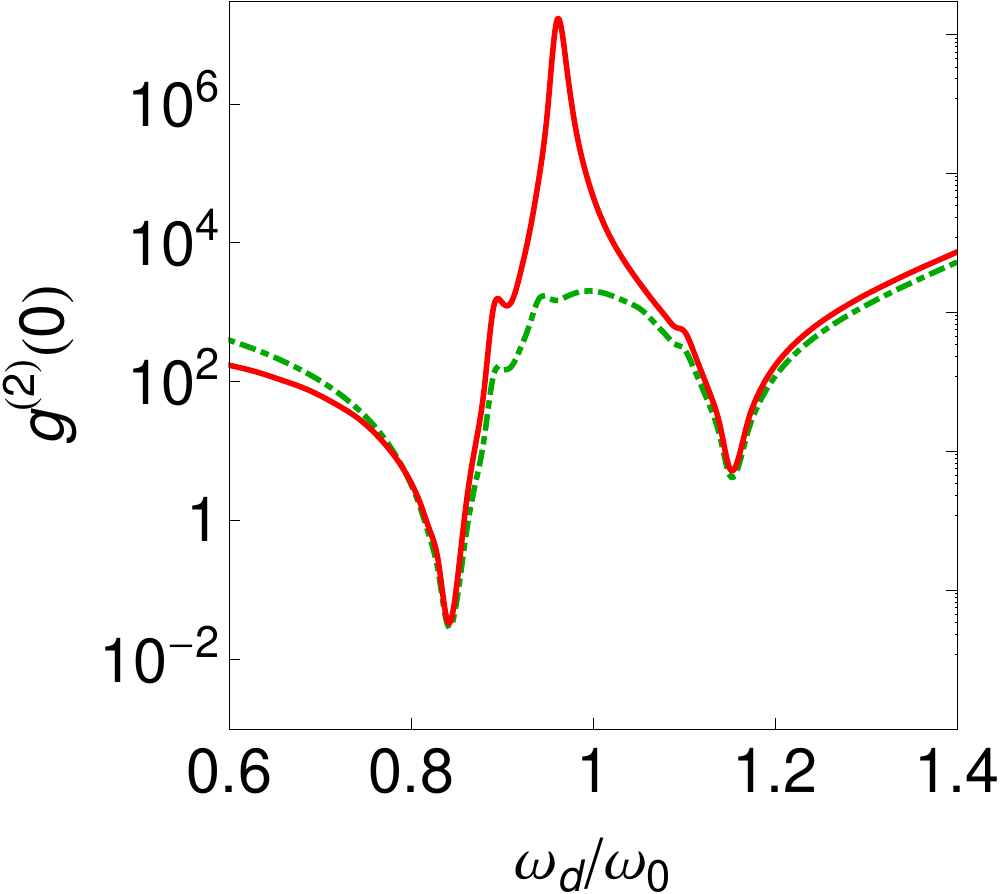}  
\caption{(color online)  Comparison between $g^{(2)}(0)$ as calculated without dephasing noise (red continuous line) and $g^{(2)}(0)$ as calculated with dephasing noise (green dash-dotted line) for $\theta = 0.93$. $\omega_{\rm x} = \omega_{\rm 0}$, $\Omega = 10^{-4}$ $\omega_{\rm 0}$, $g = 0.2$ $\omega_{\rm 0}$, $\gamma_{a} = 10^{-2}$ $\omega_{\rm 0}$ and $\gamma_{x} = 10^{-2}$ $\omega_{\rm 0}$ (red continuous line) respectively $\gamma_{\text{deph}} = \gamma_{x} = 0.5 \times 10^{-2}$ $\omega_{\rm 0}$ (green dash-dotted line).} 
\label{fig:dephased_g2}
\end{figure}

We model the dephasing with a Liouvillian superoperator $\mathcal{L}_{\text{deph}}\rho(t) = \sum_{j}\Gamma^{j}_{\text{deph}} \mathcal{D}[|j \rangle \langle j|]\rho(t)$, where $\Gamma^{j}_{\text{deph}} = \gamma_{\text{deph}} \langle j | \sigma_{\rm z} | j \rangle$ and $\mathcal{D}[\mathcal{O}]\rho = \frac{1}{2} (2 \mathcal{O}\rho\mathcal{O}^{\dagger}-\rho \mathcal{O}^{\dagger} \mathcal{O} - \mathcal{O}^{\dagger} \mathcal{O}\rho)$,
c.f. \cite{Blais}. Figure \ref{fig:dephased_g2} shows a comparison between $g^{(2)}(0)$ as calculated without dephasing noise (red continuous line) and $g^{(2)}(0)$ as calculated with dephasing noise (green dash-dotted line) for $\theta = 0.93$. $\omega_{\rm x} = \omega_{\rm 0}$, $\Omega = 10^{-4}$ $\omega_{\rm 0}$, $g = 0.2$ $\omega_{\rm 0}$, $\gamma_{a} = 10^{-2}$ $\omega_{\rm 0}$ and $\gamma_{x} = 10^{-2}$ $\omega_{\rm 0}$ (red continuous line) respectively $\gamma_{\text{deph}} = \gamma_{x} = 0.5 \times 10^{-2}$ $\omega_{\rm 0}$ (green dash-dotted line).
Hence the red continuous line in figure \ref{fig:dephased_g2} is exactly the same as the  red continuous line in figure 2a of the main text. In particular for the drive frequencies of interest close to the polariton resonances, the differences between both results are vanishingly small.

\end{document}